\documentstyle[12pt]{article}

\begin{document}
\setlength{\baselineskip}{24pt}

\title{Generalized action invariants for drift waves-zonal flow systems}
\author{A. I. Smolyakov$^{1}$ and P.H. Diamond$^{2}$ \\
$^{1}$Department of Physics and Engineering Physics, University of\\
Saskatchewan, Saskatoon, S7N5E2 Canada, \\
$^{2}$Department of Physics, University of California at San-Diego\\
La Jolla, CA 92093, USA}
\maketitle

\begin{abstract}
\setlength{\baselineskip}{24pt}
Generalized action invariants are identified for various models of drift
wave turbulence in the presence of the mean shear flow. It is shown that the
wave kinetic equation describing the interaction of the small scale
turbulence and large scale shear flow can be naturally writen in terms of
these invariants. Unlike the wave energy, which is conserved as a sum of
small- and large- scale components, the generalized action invariant is
shown to correspond to a quantity which is conserved for the small scale
component alone. This invariant can be used to construct canonical variables
leading to a different definition of the wave action ( as compared to the
case without shear flow). It is suggested that these new canonical action
variables form a natural basis for the description of the drift wave
turbulence with a mean shear flow.
\end{abstract}



The dynamics of the small scale turbulence in the presence of a mean shear
flow is a problem of a great interest for plasmas and geostrophic fluids. It
is believed that the nonlinear energy transfer from small to large length
scale component (inverse cascade \cite{hasegawa}) is a cause of a
spontaneous generation and sustainment of coherent large structures, e.g.
zonal flows in atmospheres, ocean and plasmas \cite{busse}. In the few past
years it has been suggested \cite
{rosenbluth,sydora,lin,hammett,dimits,waltz,diamondkim,iaea98} that the
large scale flow band structures (zonal flows) play an important role in
regulating and suppressing the anomalous transport in magnetic confinement
systems.

In the simplest form, the generation of plasma flow by turbulence can be
described by the energy conservation relation (Poynting theorem) averaged
over small scale fluctuations \cite{diamondkim}. A generalization of this
approach is a WKB type wave kinetic equation for the quanta density of small
scale fluctuations that is conserved along the rays. This method was
originally proposed in Ref. 11 to describe the interaction of high frequency
plasmons (Langmuir waves) with low frequency ion sound perturbations. It is
widely used also in general fluid dynamics \cite{andrews}.

In studies of drift wave dynamics, it has been naturally assumed \cite
{mattor94,brizard} that the relevant quantity that is conserved in the
presence of slow variations is the drift-wave action density. It is well
known \cite{sagdeev}, that the standard wave action variables $C_{k}$
associated with the number of wave quanta $n_{k}$, $n_{k}=\left|
C_{k}\right| ^{2}=E_{k}/\omega _{k}$, where $E_{k}$ is the wave energy, and $%
\omega _{k}$ is the wave frequency, is a basis for Hamiltonian form of the
wave-wave interaction equations. It has been noted in Refs. 16,17 that the
normal variables used to describe self-interaction between small scale
fluctuations without the shear flow are modified by the flow and may not be
suitable for a system with a mean flow. Thus, in the presence of a shear
flow a new form of canonical variables and associated action invariant have
to be identified. On other hand, it has been pointed\cite{lebedev} that the
conserved action-like quantity (pseudo-action) is different from the usual
definition of the wave action defined as the ratio of the wave energy to the
wave frequency. The latter definition is also fails when there are no
oscillating eigenmodes such as in ideal fluid, so that an alternative
definition of the action-like integral is required\cite{nazarenko}.

It is important to realize that the natural form of the three-wave
interaction equations for the drift-waves does not have Hamiltonian
structure \cite{zakharov}.These equations can be transformed, however, to a
Hamiltonian form via an asymptotic variable transformation. Such a
tranformation yielding a Hamiltonian form for the drift and Rossby waves has
been found in Refs. 20,21. There are several possible forms for such a
transformation. In Refs. 17,20,21 it is based on the conserved energy
integral that leads to the standard definition of the wave action. For
drift-wave+zonal flow systems small scales are modulated by larger scale
shear flows so that energy in the small scale component is not conserved.
Thus, the canonical Hamiltonian variables constructed from energy
conservation are not suitable for description of the drift waves in the
presence of a mean flow.

In this paper, we derive the WKB type wave kinetic equation that describes
the conservation (along the rays) of an action like invariant of the drift
wave turbulence with slowly varying parameters due to the mean sheared flow.
We demonstrate that the relevant action-like integral corresponds to the
quantity conserved for the small scale component alone. We show that the
structure of the action integral is determined by the structure of the
matrix element describing the interaction of the small scale and large scale
component. We discuss how the canonical variables corresponding to such a
pseudo-action invariant can be constructed.

The scale separation between the small scale turbulence and the large scale
motions is an essential property of drift-wave+zonal flow systems that is
commonly used \cite{andrews,manin,balk,lebedev,nazarenko,muhm,dyachenko} to
simplify the analysis. Though, the scale separation is often observed
experimentally and in computer simulation, it may be less pronounced in
other cases\cite{hahm}. In our present paper, we substantially rely on the
multiscale expansion, so our results are valid, strictly speaking, only in
the case when there is such a scale separation. More general approach
avoiding the scale separation assumption, namely the renormalization group,
is possible \cite{rg}, but it is beyond the scope of the present paper.

We consider a generic case of the drift wave equation in the form 
\begin{equation}
\frac{\partial \phi _{k}}{\partial t}+i\omega _{k}\phi _{k}+\int
d^{2}pL_{p,k-p}\phi _{p}\phi _{k-p}=0,  \label{gen}
\end{equation}
where $\omega _{k}=\omega (k)$ is the frequency of the linear mode with a
wavector $k$, and may include an imaginary part corresponding to the wave
grow and decay.

In the spirit of the scale separation we represent the field into the
large-scale $\phi _{k}^{<}$ and small-scale $\phi _{k}^{>}$ components; $%
\phi _{k}^{<}=0$ outside a shell $\mid {\bf k}\mid <\varepsilon \ll 1$, $%
\phi _{k}^{>}=0$ for $\mid {\bf k}\mid <\varepsilon .$

Assuming that the self-interaction of small-scale fields is small compared
to the interaction with the mean flow\cite{balk} we write from (\ref{gen})
the following equation for the small-scale fluctuations 
\begin{equation}
\frac{\partial \phi _{k}^{>}}{\partial t}+i\omega _{k}\phi _{k}^{>}+\int
d^{2}pL_{p,k-p}\phi _{p}^{<}\phi _{k-p}^{>}=0.  \label{g1}
\end{equation}
To derive the equation for the evolution of the wave spectrum we multiply
equation (\ref{g1}) by $\phi _{k^{^{\prime }}}^{>}$ and then add it with a
similar equation obtained by reversing $k$ and $k^{^{\prime }}$, yielding 
\begin{equation}
\frac{\partial }{\partial t}\left( \phi _{k}^{>}\phi _{k^{^{\prime
}}}^{>}\right) +i\left( \omega _{k}+\omega _{k^{^{\prime }}}\right) \ \phi
_{k}^{>}\phi _{k^{^{\prime }}}^{>}+\phi _{k^{^{\prime }}}^{>}\int
d^{2}pL_{p,k-p}\phi _{p}^{<}\phi _{k-p}^{>}+\phi _{k}^{>}\int
d^{2}pL_{p,k^{^{\prime }}-p}\phi _{p}^{<}\phi _{k^{^{\prime }}-p}^{>}=0.
\label{g2}
\end{equation}

The small-scale turbulence is described by the spectral function (Wigner
function) $I_{k}({\bf x},t),$ defined as follows 
\begin{equation}
\int d^{2}q\left\langle \phi _{-k+q}^{>}\phi _{k}^{>}\right\rangle \exp (i%
{\bf q\cdot x})=I_{k}({\bf x},t).  \label{ik}
\end{equation}

The slow time and spatial dependence in $I_{k}({\bf x},t)$ corresponds to
modulations with a ``slow'' wavevector, ${\bf q\ll k}$. Angle brackets in (%
\ref{ik}) stand for ensemble average, which is equivalent to a time average
with appropriate ergodic assumptions.

The equation for $I_{k}({\bf x},t)$ is derived from (\ref{g2}) by averaging
it over fast scales and by taking the Fourier transform over the slow
variable ${\bf x}$. Setting ${\bf k}^{^{\prime }}=-{\bf k}+{\bf q}$ and
applying the operator $\int d^{2}q\exp (i{\bf q\cdot x})$ we obtain

\begin{equation}
\frac{\partial }{\partial t}I_{k}({\bf x},t)+i\int d^{2}q\exp (i{\bf q\cdot x%
})\left( \omega _{k}+\omega _{-k+q}\right) \left\langle \phi _{k}^{>}\phi
_{-k+q}^{>}\right\rangle \ +S_{1}+S_{2}=0,  \label{nk}
\end{equation}

\begin{equation}
S_{1}=\int \int d^{2}pd^{2}q\exp (i{\bf q\cdot x})\left\langle \phi
_{-k+q}^{>}\phi _{k-p}^{>}\right\rangle L_{p,k-p}\phi _{p}^{<},
\end{equation}
\begin{equation}
S_{2}=\int \int d^{2}pd^{2}q\exp (i{\bf q\cdot x})\left\langle \phi
_{-k+q-p}^{>}\phi _{k}^{>}\right\rangle L_{p,-k+q-p}\phi _{p}^{<}.
\end{equation}
The second term in (\ref{g2}) gives 
\begin{equation}
i\int d^{2}q\exp (i{\bf q\cdot x})\left( \omega _{k}+\omega _{-k+q}\right)
\left\langle \phi _{k}^{>}\phi _{-k+q}^{>}\right\rangle =\frac{\partial
\omega _{k}}{\partial {\bf k}}\cdot \frac{\partial }{\partial {\bf x}}I_{k}(%
{\bf x},t)-2\gamma _{k}I_{k},  \label{ikl}
\end{equation}
where $\gamma _{k}$ is the linear growth rate, and only the real part of the
frequency is presumed for $\omega _{k}\,$on the right hand side of this
equation.

The ensemble average in $S_{1}$ can be transformed by using the inverse of (%
\ref{ik}) 
\begin{equation}
\left\langle \phi _{-k+q}^{>}\phi _{k-p}^{>}\right\rangle =\left\langle \phi
_{k-p}^{>}\phi _{-(k-p)+q-p}^{>}\right\rangle =\int d^{2}x^{^{\prime
}}I_{k-p}(x^{^{\prime }})\exp (-i({\bf q}-{\bf p})\cdot {\bf x}^{^{\prime
}}).  \label{in1}
\end{equation}
By using (\ref{in1}) and expanding in ${\bf p}\ll {\bf k}$ the expression
for $S_{1}$ is transformed to 
\begin{equation}
S_{1}=\int d^{2}p\exp (i{\bf p\cdot x})L_{p,k-p}\left( I_{k}({\bf x})-{\bf p}%
\cdot \frac{\partial I_{k}({\bf x})}{\partial {\bf k}}\right) \phi _{p}^{<}{%
\ .}\ {\ }  \label{s1}
\end{equation}
Similarly, by using the identity analogous to (\ref{in1}) and expanding the
interaction coefficient $L_{p,k-p}$ in ${\bf p}\ll {\bf k,}$ we transform $%
S_{2}$ to the form 
\begin{eqnarray}
S_{2} &=&\int \int d^{2}pd^{2}q\exp (i{\bf q\cdot x})\left( L_{p,-k}+({\bf q}%
-{\bf p)}\cdot \frac{\partial L_{p,-k}}{\partial (-{\bf k)}}\right) \phi
_{p}^{<}  \nonumber \\
&&\times \int d^{2}x^{^{\prime }}{\ }\exp (-i({\bf q-p}){\bf \cdot x}%
^{^{\prime }})I_{k}(x^{^{\prime }})  \nonumber \\
{} &=&{}I_{k}(x)\int d^{2}p\exp (i{\bf p\cdot x})L_{p,-k}\phi _{p}^{<}-i\int
d^{2}p\exp (i{\bf p\cdot x})\frac{\partial L_{p,-k}}{\partial (-{\bf k)}}%
\cdot \frac{\partial I_{k}}{\partial {\bf x}}\phi _{p}^{<}\ .\ {\ }
\label{s2}
\end{eqnarray}

Equations (\ref{nk}-\ref{s2}) define a particular form of the transport
equation for $I_{k}({\bf x},t)$ for a given interaction coefficient $%
L_{k,k^{^{\prime }}}$.

In this paper, we consider two different models for drift waves in a
magnetized plasma: the standard Hasegawa-Mima equation and a slab-like model
for drift waves in a sheared magnetic field. The latter is similar to the
standard Hasegawa-Mima equation with a modified plasma response to the slow
modulations of the electrostatic potential. Such slow modes correspond to $%
k_{\Vert }\rightarrow 0$, so that the slow part of the potential does not
follow Boltzmann distribution. [Note that zonal flows\cite{iaea98} ($m=n=0$)
are such slow modes with $k_{\Vert }=0$.] As a result, the convective term
appears in the lowest order, contrary to the case of the Hasegawa-Mima
equation where such term is due to the polarization drift. Appropriate
equation for the drift wave dynamics in presence of a mean flow (neglecting
the self-interaction) has the form \cite{mattor94} 
\begin{equation}
\left( \frac{\partial }{\partial t}+{\bf V}_{0}\cdot \nabla \right) \frac{e%
\widetilde{\phi }}{T_{e}}+{\bf V}_{*}\cdot \nabla \frac{e\widetilde{\phi }}{%
T_{e}}-\rho _{s}^{2}\left( \frac{\partial }{\partial t}+{\bf V}_{0}\cdot
\nabla \right) \nabla _{\bot }^{2}\frac{e\widetilde{\phi }}{T_{e}}=0.
\label{dwr}
\end{equation}
where ${\bf V}_{0}=c{\bf b\times }\nabla \overline{\phi }/B_{0}$ is the mean
flow velocity. This equation can be written in the form (\ref{g1}) with $%
\omega _{k}={\bf k\cdot V}_{*}/(1+k^{2}\rho _{s}^{2})$ and 
\begin{equation}
L_{k_{1},k_{2}}=-\frac{c}{B_{0}}\frac{{\bf b\cdot k}_{1}\times {\bf k}_{2}}{%
1+({\bf k}_{1}+{\bf k}_{2})^{2}\rho _{s}^{2}}\left( 1+k_{2}^{2}\rho
_{s}^{2}\right) .  \label{c1}
\end{equation}

From (\ref{nk}-\ref{s2}) and (\ref{c1}) we obtain 
\begin{equation}
\frac{\partial }{\partial t}I_{k}({\bf x},t)+\frac{\partial }{\partial {\bf k%
}}\left( \omega _{k}+{\bf k\cdot V}_{0}\right) \cdot \frac{\partial I_{k}}{%
\partial {\bf x}}-\frac{\partial }{\partial {\bf x}}\left( \frac{{\bf k\cdot
V}_{0}}{(1+k^{2}\rho _{s}^{2})^{2}}\right) \cdot \frac{\partial }{\partial 
{\bf k}}I_{k}(1+k^{2}\rho ^{2})^{2}=0.  \label{n1}
\end{equation}
This equation can be written in the form of a conservation law for the
invariant $N_{k}=I_{k}(1+k^{2}\rho ^{2})^{2},$ 
\begin{equation}
\frac{\partial }{\partial t}N_{k}({\bf x},t)+\frac{\partial }{\partial {\bf k%
}}\left( \omega _{k}+{\bf k\cdot V}_{0}\right) \cdot \frac{\partial N_{k}}{%
\partial {\bf x}}-\frac{\partial }{\partial {\bf x}}\left( {\bf k\cdot V}%
_{0}\right) \cdot \frac{\partial }{\partial {\bf k}}N_{k}=0.  \label{nk1}
\end{equation}

By direct evaluation from (\ref{dwr}), it can be easily shown that the
quantity 
\begin{equation}
N=\int d^{2}k\left( \widetilde{\phi }^{2}+2\rho _{s}^{2}(\nabla _{\bot }%
\widetilde{\phi })^{2}+\rho _{s}^{4}(\nabla _{\bot }^{2}\widetilde{\phi }%
)^{2}\right) ,  \label{nr}
\end{equation}
corresponding to $N_{k}$ in (17), is conserved as an integral over the
small-scale part of the spectrum. In (\ref{nr}) $\widetilde{\phi }$ is the
normalized potential of the small scale fluctuations. This property
distinguishes $N_{k}$ from any other combination of the energy and enstrophy
which are conserved only as a sum of contributions from the small and long
scale parts of the spectrum\cite{muhm}.

A different expression for the action-like invariant is obtained for the
standard Hasegawa-Mima (H.M.) model with a mean flow 
\begin{equation}
\frac{\partial }{\partial t}\left( \frac{e\widetilde{\phi }}{T_{e}}-\rho
_{s}^{2}\nabla _{\bot }^{2}\frac{e\widetilde{\phi }}{T_{e}}\right) +{\bf V}%
_{*}\cdot \nabla \frac{e\widetilde{\phi }}{T_{e}}-\rho _{s}^{2}({\bf V}%
_{0}\cdot \nabla )\nabla _{\bot }^{2}\frac{e\widetilde{\phi }}{T_{e}}=0.
\label{hm}
\end{equation}

The appropriate interaction coefficient is 
\begin{equation}
L_{k_{1},k_{2}}=-\frac{c}{2B_{0}}\rho _{s}^{2}\frac{{\bf b\cdot k}_{1}\times 
{\bf k}_{2}}{1+({\bf k}_{1}+{\bf k}_{2})^{2}\rho _{s}^{2}}\left(
k_{2}^{2}-k_{1}^{2}\right) .  \label{c2}
\end{equation}

In this case, from (5-11) and (\ref{c2}) the transport equation for $I_{k}$
takes the form 
\begin{eqnarray}
&&\frac{\partial }{\partial t}I_{k}+\frac{\partial }{\partial {\bf k}}\left(
\omega _{k}+\frac{{\bf k\cdot V}_{0}}{1+k^{2}\rho _{s}^{2}}k^{2}\rho
_{s}^{2}\right) \cdot \frac{\partial }{\partial {\bf x}}I_{k}  \nonumber \\
&&-\frac{\partial }{\partial {\bf x}}\left( \frac{{\bf k\cdot V}_{0}}{%
(1+k^{2}\rho _{s}^{2})^{2}}\right) \cdot \frac{\partial }{\partial {\bf k}}%
k^{2}\rho _{s}^{2}(1+k^{2}\rho _{s}^{2})I_{k}=0.  \label{nhm}
\end{eqnarray}
Obviously, this equation can be written in the form of the conservation law
for the invariant $N_{k}=I_{k}k^{2}\rho _{s}^{2}(1+k^{2}\rho _{s}^{2}),$\cite
{lebedev,muhm,dyachenko} 
\begin{equation}
\frac{\partial }{\partial t}N_{k}+\frac{\partial }{\partial {\bf k}}\left(
\omega _{k}+\frac{{\bf k\cdot V}_{0}}{1+k^{2}\rho _{s}^{2}}k^{2}\rho
^{2}\right) \cdot \frac{\partial }{\partial {\bf x}}N_{k}-\frac{\partial }{%
\partial {\bf x}}\left( \frac{{\bf k\cdot V}_{0}}{(1+k^{2}\rho _{s}^{2})}%
k^{2}\rho _{s}^{2}\right) \cdot \frac{\partial }{\partial {\bf k}}N_{k}=0.
\label{whm}
\end{equation}

Similarly to the previous case, the invariant $N_k$ corresponds to the
integral of (\ref{hm}) conserved for the small scale component alone

\begin{equation}
N=\int d^{2}k{\ }\rho _{s}^{2}\left( (\nabla _{\bot }\widetilde{\phi }%
)^{2}+\rho _{s}^{2}(\nabla _{\bot }^{2}\widetilde{\phi })^{2}\right) ,
\label{nhm2}
\end{equation}
Note that both invariants (\ref{nr}) and (\ref{nhm2}) are different from
standard definition of the wave action \cite{mattor94,brizard}. The
difference between two forms of the action-like invariant (Eq. (\ref{nr})
and (\ref{nhm2})) is due to a different form of the coupling matrix (Eq. (%
\ref{c1}) and Eq.(\ref{c2})) describing interaction of the small and large
scale components.

The procedure that we have described above can also be used to derive the
action-like invariant for the two-dimensional motion of an incompressible
fluid. In the latter case, there are no oscillating modes so that the
standard definition of the action as a ratio of the wave energy to wave
frequency is not applicable. The 2-D Euler equation has a form 
\begin{equation}
\partial \nabla _{\bot }^{2}\phi +{\bf V}_{0}\cdot \nabla \nabla _{\bot
}^{2}\phi =0,  \label{e}
\end{equation}
where ${\bf V}_{0}$ is the velocity due to the mean flow. This equation can
be written in the form (\ref{gen}) with $\omega _{k}=0$ and the interaction
coefficient 
\begin{equation}
L_{k_{1},k_{2}}=-\frac{{\bf b\cdot k}_{1}\times {\bf k}_{2}}{({\bf k}_{1}+%
{\bf k}_{2})^{2}}k_{2}^{2}.  \label{ce}
\end{equation}

Using equations (5-11) and (\ref{ce}) we obtain the wave kinetic equation 
\begin{equation}
\frac{\partial }{\partial t}N_{k}({\bf x},t)+\frac{\partial }{\partial {\bf k%
}}\left( {\bf k\cdot V}_{0}\right) \cdot \frac{\partial N_{k}}{\partial {\bf %
x}}-\frac{\partial }{\partial {\bf x}}\left( {{\bf k\cdot V}_{0}}{}\right)
\cdot \frac{\partial }{\partial {\bf k}}N_{k}=0,
\end{equation}
where the wave-action $N_{k}=k^{4}I_{k}$ \cite{nazarenko}.

We summarize generalized wave action integrals for different models in the
Table I. Note that the standard expression for the drift wave action defined
as the ratio of the wave energy to the wave frequency is \cite
{mattor94,brizard} 
\begin{equation}
n_{k}=\left| a_{k}\right| ^{2}=\frac{(1+\rho _{s}^{2}k_{\bot }^{2})^{2}}{%
\omega _{*}}\left| \phi _{k}\right| ^{2}=\frac{E_{k}}{\omega _{k}},
\label{a}
\end{equation}
where $\omega _{*}=k_{\theta }V_{*}$. Expression (\ref{a}) should be
compared with the first two lines in the Table. It is interesting to note
that generalized action invariant given by Eq. (\ref{nr}) coincides with the
standard definition of the wave action (\ref{a}) for the poloidally
symmetric case when the poloidal wave vector $k_{\theta }$ is constant ($%
k_{\theta }=const$).

Next we consider the self-interaction between small scales in the presence
of the shear flow and outline how the pseudo-action invariants can be used
to construct the canonical variables for the latter case. For illustration,
we consider the case of Hasegawa-Mima equation (\ref{hm}). We restore the
self-interaction term given by $W_{k,k_{1},k_{2}}$ 
\begin{equation}
\frac{\partial \phi _{k}}{\partial t}+i\omega _{k}\phi _{k}=\int
d^{2}k_{1}d^{2}k_{2}W_{k,k_{1},k_{2}}\delta (k-k_{1}-k_{2})\phi _{k_{1}}\phi
_{k_{2}},  \label{dphik}
\end{equation}
\begin{equation}
W_{k,k_{1},k_{2}}=-\frac{c}{2B_{0}}\rho _{s}^{2}\frac{{\bf b\cdot k}%
_{1}\times {\bf k}_{2}}{1+{\bf k}^{2}\rho _{s}^{2}}\left(
k_{2}^{2}-k_{1}^{2}\right) .
\end{equation}
This natural form of the three-wave interaction does not have standard
Hamiltonian structure. This is reflected in the interaction coefficients $%
W_{k,k_{1},k_{2}}$ which do not have the required symmetry properties \cite
{sagdeev}. The only symmetries in $W_{k,k_{1},k_{2}}$ are of the type $%
W_{-k,-k_{1},-k_{2}}^{*}=W_{k,k_{1},k_{2}}=W_{-k,k_{1},k_{2}}.$
Transformation of (\ref{dphik}) to normal canonical variables $a_{k}$ was
given in Refs. 19,20 (see also Ref. 16). It has the form \cite{balk} 
\begin{equation}
a_{k}=g_{k}\phi _{k}+\int d^{2}k_{1}d^{2}k_{2}G_{k,k_{1},k_{2}}\delta
(k-k_{1}-k_{2})\phi _{k_{1}}\phi _{k_{2}}.
\end{equation}
In new variables the interaction coefficients $V_{k,k_{1},k_{2}}$ are 
\begin{equation}
V_{k,k_{1},k_{2}}=\frac{1}{3g_{k_{1}}g_{k_{2}}g_{k}}\left( \left|
g_{k}\right| ^{2}W_{k,k_{1},k_{2}}+\left| g_{k_{1}}\right|
^{2}W_{k1,k,k_{2}}+\left| g_{k_{2}}\right| ^{2}W_{k_{2},k_{1},k}\right)
\end{equation}
These interaction coefficients $V_{k,k_{1},k_{2}}$ now have all symmetries
required for Hamiltonian systems. The function $g_{k}$ can be chosen in a
variety of ways. The standard approach \cite{balk,zakharov,monin} is to
chose $g_{k}$ so that the energy in canonical variables takes the form $%
E=\int d^{2}k\omega _{k}a_{k}a_{-k}.$ Comparing it with the energy integral $%
E=\int d^{2}k\left( \widetilde{\phi }^{2}+\rho _{s}^{2}(\nabla _{\bot }%
\widetilde{\phi })^{2})\right) ,$we find \cite{balk} $g_{k}=(1+\rho
_{s}^{2}k_{\bot }^{2})/(k_{y})^{1/2}$. This gives a standard expression for
the wave action (\ref{a}).

As discussed above, for the drift waves-zonal flow system the energy in the
small scale component is not conserved, bur rather the total energy of drift
waves + large scale zonal flows is constant. For this reason, the energy
integral of the small scale component can not be used for introduction of
canonical variables for self interaction of the small scale fluctuations.
Contrary to the energy, the integrals $N_{k}$ are conserved for small scale
component. Choosing the function $g_{k}$ such as that the invariants (\ref
{nr}) or (\ref{nhm2}) are in the form $N_{k}=\int d^{2}ka_{k}a_{-k},$we
obtain $N_{k}$ as canonical variables for drift waves in the presence of the
mean shear flow. This automatically means that these invariants have a
meaning of the generalized wave action invariant. Then, to account for the
self-interactions in the presence of the background shear flow, the wave
kinetic equation (Eq. (\ref{nk1}) or (\ref{whm})) should be modified with a
source term $J_{k}$ in the standard form\cite{sagdeev} 
\begin{equation}
J_{k}=4\pi \int d^{2}k_{1}d^{2}k_{2}\times \left| V_{k,k_{1},k_{2}}\right|
^{2}\left( N_{k_{1}}N_{k_{2}}-N_{k}N_{k_{1}}-N_{k}N_{k}\right) \delta
(k-k_{1}-k_{2}).
\end{equation}

We have formulated a wave kinetic equation and determined a structure of an
appropriate adiabatic invariant for small scale turbulence in the presence
of a mean flow. We have shown that the form of the matrix coefficient for
the nonlocal coupling of the small scale fluctuations to the mean flow is
crucial for the form of the adiabatic invariant. We have obtained adiabatic
invariant $N_{k}=I_{k}k^{2}\rho _{s}^{2}(1+k^{2}\rho _{s}^{2})$ for the
drift wave turbulence described by the Hasegawa-Mima equation and isomorphic
Charney-Obukhov equation for Rossby waves; and the invariant $%
N_{k}=I_{k}(1+k^{2}\rho _{s}^{2})^{2}$ for the drift wave type turbulence in
tokamaks such as TITG driven modes. [Note that the latter invariant reduces
to the standard form \cite{mattor94,brizard} for $k_{\theta }=const$.] The
pseudo-action invariants appear in the wave kinetic equation and correspond
to the quantities that are conserved as integrals over the small scale part
of the spectrum alone. This specific conservation property makes them
suitable as canonical Hamiltonian variables for small scale turbulence in
the presence of the shear flow. The wave action invariants and the kinetic
equation derived here can be used to investigate nonlinear dynamics of drift
waves and zonal flow in a tokamak. The method used in our work can be
applied to derive generalized invariants for other models including the
Rossby type waves in geostrophic fluids \cite{andrews}.

This research was supported by Natural Sciences and Engineering Research of 
Canada and U.S. Department of Energy Grant No. FG03-88ER53275. P.D. would
like to thank V.B. Lebedev, M.N. Rosenbluth and F.L. Hinton for helpful
discussions.

\newpage

Table I: Generalized action invariants for different models  \vspace{.3cm}

\begin{tabular}{cc}
\hline\hline
Model & Expression for the wave action \\ \hline
Drift waves in a sheared field, Eq.(\ref{dwr}) & $I_{k}(1+k^{2}\rho
_{s}^{2})^{2}$ \\ 
Standard drift wave model, Eq.(\ref{hm}) & $I_{k}k^{2}\rho
_{s}^{2}(1+k^{2}\rho _{s}^{2})$ \\ 
2D Euler equation, Eq. (\ref{e}) & $I_{k}k^{4}$ \\ \hline\hline
\end{tabular}

\end{document}